# Device Perspective for Black Phosphorus Field-Effect Transistors: Contact Resistance, Ambipolar and Scaling

Yuchen Du, Han Liu, Yexin Deng and Peide D. Ye[*]

*School of Electrical and Computer Engineering and Birck Nanotechnology Center,*

*Purdue University, West Lafayette, Indiana 47907, United States*

Correspondence to: yep@purdue.edu

Abstract

Although monolayer black phosphorus (BP) or phosphorene has been successfully exfoliated and its optical properties have been explored, most of electrical performance of the devices is demonstrated on few-layer phosphorene and ultra-thin BP films. In this paper, we study the channel length scaling of ultra-thin BP field-effect transistors (FETs), and discuss a scheme for using various contact metals to change transistor characteristics. Through studying transistor behaviors with various channel lengths, the contact resistance can be extracted from the transfer length method (TLM). With different contact metals, we find out that the metal/BP interface has different Schottky barrier heights, leading to a significant difference in contact resistance, which is quite different from previous studies of transition metal dichalcogenides (TMDs) such as $MoS_2$ where Fermi-level is strongly pinned near conduction band edge at metal/$MoS_2$ interface. The nature of BP transistors are Schottky barrier FETs, where the on and off states are controlled by tuning the Schottky barriers at the two contacts. We also observe the ambipolar characteristics of BP transistors with enhanced n-type drain current and demonstrate that the p-type carriers can be easily shifted to n-type or vice versus by controlling the gate bias and drain bias, showing the potential to realize BP CMOS logic circuits.

Low dimensional materials are getting more and more interests in condensed-matter physics, materials, and device communities. They provide a new class of materials that are atomic-layer thick. Graphene, boron nitride, and TMDs provide the ideal metal, insulator and semiconductors as three basic building blocks for any device applications.[1-15] Transistors built on monolayer or few-layer $MoS_2$, the most studied TMDs, are the optimal forms of ultra-thin body FETs with an ideal structure to immune the short channel effects.[9,13] Also, the comparatively heavier effective mass of the $MoS_2$ allows transistors to have lower direct source–drain leakage current, increased drive current, and enhanced transconductance when benchmarked against the ultra-thin body Si transistors at their scaling limit.[14] However, the carrier mobility of $MoS_2$ is much lower than that of graphene or traditional semiconductors.[16] Due to the presence of S vacancies and strong Fermi level pinning near the conduction band,[17] $MoS_2$ transistors monotonously show n-type FET characteristics even when using high work functional contact metals.[18]

Quest for p-type 2D materials is needed to build up energy-efficient 2D electronic or optoelectronic devices such as CMOS, tunneling FETs, photo-detectors, and solar cells beyond 2D Schottky-barrier FETs. In this paper, we take a deep insight into the metal/BP interface from the device aspects by applying different high work-function metals (Ni and Pd) on BP field-effect transistors with channel length scaling from 3 μm down to 100 nm. BP, the bulk counterpart of phosphorene, is a stable phosphorus allotrope at room temperature.[19,20] BP has the layered structure in which individual atomic layers are stacked together by Van der Waals interactions.[21-23] The fundamental properties of bulk BP had been extensively studied in the past, indicating the bulk crystal of BP is a semiconducting material with a direct bandgap of 0.3 eV,[19,21,23] and hole mobility in black phosphorus can reach up to $10^4$ cm$^2$/Vs.[24] The new wave of interests in BP is inspired by graphene research, attempt to realize monolayers of BP coined as phosphorene. Although phosphorene has been successfully exfoliated and its optical properties have been explored,[25] most of electrical performance of the devices is demonstrated on few-layer phosphorene and ultra-thin BP films.[25-29] This is because phosphorene is less stable in air in particular with the presence of both oxygen and water.[30] An effective passivation technique on phosphorene is needed for practical applications. Here, we focus on ultra-thin BP films since the large size of the flakes (tens of μm) is needed so that we can fabricate numerous FETs with two different metals on the same flake for the fair comparison as shown in Figure 1(a). Fortunately, few-layer or ultra-thin 2D films are more favorable for device demonstrations because they can deliver larger current than monolayer in general. We determine the contact resistance of Ni and Pd on BP of 1.75±0.06 Ω·mm versus 3.15±0.15 Ω·mm at high electrostatic doping limit. Furthermore, the characteristics of BP transistors with Ni contacts can be switched from the p-type behavior into pronounced n-type once scaled down to deep sub-micron channel length. The results all confirm the nature of Schottky barrier FETs for BP transistors with its smaller bandgap.

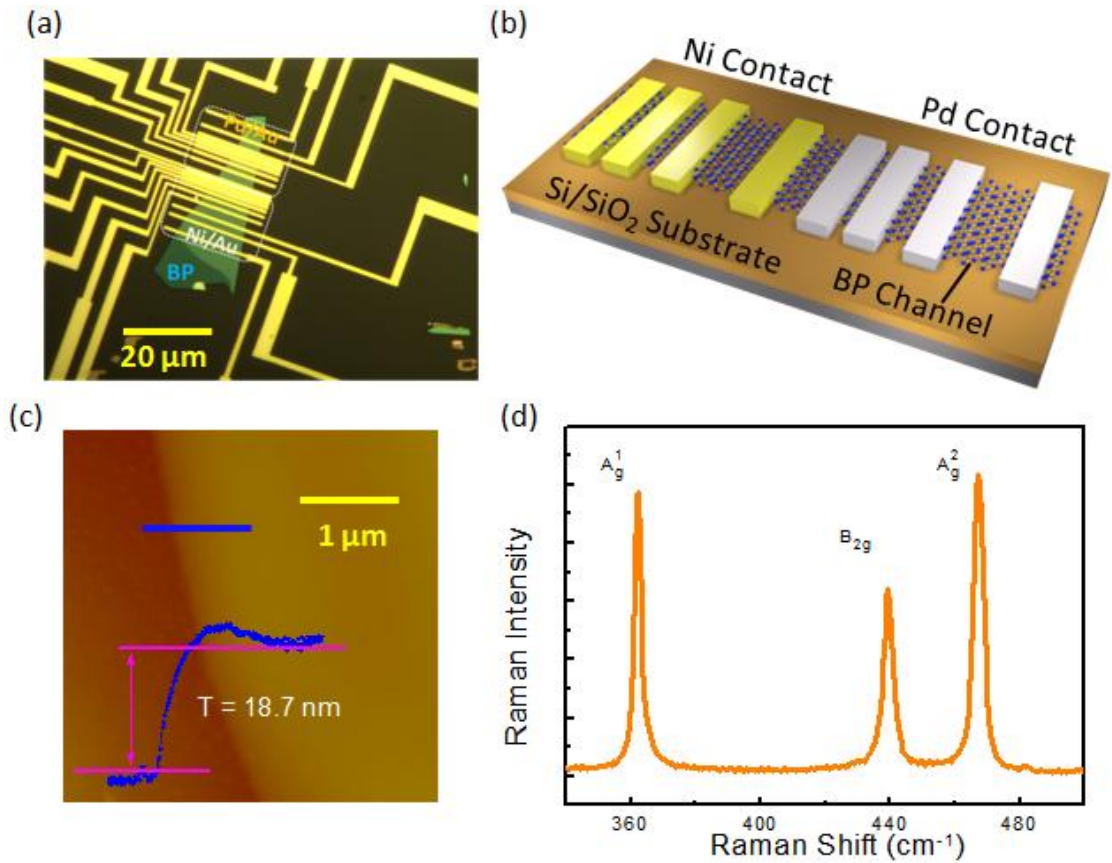

Figure 1: (a) Optical image of the ultra-thin BP FETs with Ni/Au and Pd/Au contacts on the same BP flake. (b) Schematic view of device configurations. A $p^{++}$ silicon wafer capped with 90 nm $SiO_2$ was used as the global gate and gate dielectric, respectively. Ultra-thin BP films were exfoliated from bulk BP crystals. Metal contacts with various channel lengths from 3 μm down to 100 nm for both Ni/Au and Pd/Au were evaporated onto the same BP flake. (c) Atomic force microscopy image of this large size BP film with a measured thickness of 18.7 nm. (d) Raman spectra of the same BP film.

**Results/Discussion**

In our experiments, a large size of ~40 μm x ~20 μm BP flake was exfoliated from bulk BP crystal using the scotch tape method, and then transferred to a heavily doped silicon substrate with a 90-nm $SiO_2$ capping layer. Prior to device fabrication, the ultra-thin BP film was soaked in acetone for 2 hours to remove tape residues. Contact bars with lengths of 0.5 μm were defined by the e-beam lithography. The channel lengths are designed to be 3 μm, 2 μm, 1.5 μm, 1 μm, 500 nm, 200 nm, and 100 nm. Both 30/50 nm Ni/Au and 30/70 nm Pd/Au, deposited via e-beam evaporation, were used as contact metals on the same BP flake for a fair comparison. The optical image and schematic view of the ultra-thin BP FETs are illustrated in Figure 1(a) and 1(b), respectively. The thickness of BP for this study is 18.7 nm shown in Figure 1(c), measured by the atomic force microscopy (AFM). The Raman spectrum of the channel area of BP FET is depicted in Figure 1(d). The peaks at 362 cm$^{-1}$, 439 cm$^{-1}$, and 467 cm$^{-1}$ are due to vibrations of the crystalline lattice of the BP and they match the Raman shifts attributed to the $A^1_g$, $B^2_g$ and $A^2_g$ phonon modes observed in bulk black phosphorus, and recent Raman studies of phosphorene.[25,27,28]

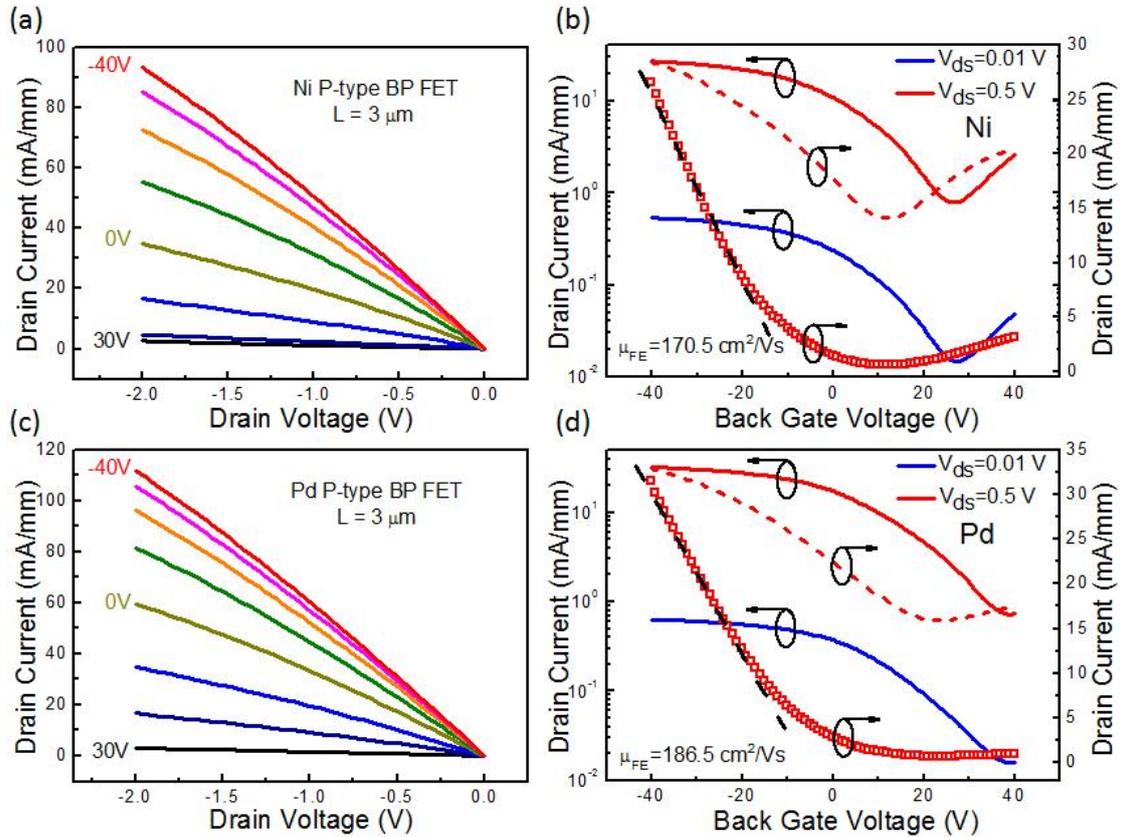

Figure 2: Device performance of p-type BP transistors for both Ni and Pd contacts.

Output (a) and transfer (b) characteristics of Ni contact BP transistor with a channel length of 3 μm. Output (c) and transfer (d) characteristics of Pd contact BP transistor with a channel length of 3 μm. The arrow directions in transfer curve stand for back gate sweep directions. Gate bias sweeps from -40 V to 40 V for Ni and Pd transfer characteristics are shown in both semi-logarithmic scale (red dashed line) and linear scale (red square).

*I-V* output characteristic of an 18.7 nm thick, 3 μm long, BP FET with Ni contact is shown in Figure 2 (a) with a back gate sweep from -30 V to 40 V. On-state current increases as the back gate sweeps from positive voltage to negative voltage, which is a clear signature of p-type transistor behavior. The drain current varies linearly with small source/drain biases, demonstrating an Ohmic-like contact resistance at the metal/BP interface with a small Schottky barrier. On-state drain current of 93.3 mA/mm for the 3 μm channel length in Ni contact device is observed with $V_{bg}$ = -40 V, and $V_{ds}$ = -2 V. The transfer curve of Ni device measured at 0.01 V and 0.5 V drain bias is shown in Figure 2(b). The transistor shows a clear switching behavior with a low current on/off ratio of ~$10^2$, due to the thick flake and low bandgap of BP.[25-27] Inspecting the transfer behavior of the p-type BP transistor, we can estimate the field-effect mobility by referring to the simple square law theory. The extrinsic field-effect mobility, $\mu_{FE}$, is 170.5 cm$^2$/Vs, as calculated from $\mu_{FE} = g_m L/(C_{ox} V_{ds} W)$, where $C_{ox}$ is the capacitance of the gate oxide, W and L are the channel width and length, $V_{ds}$ is the drain bias, and $g_m$ is the peak transconductance extracted from linear scale of transfer behaviors. The hysteresis from different back gate sweeps is due to the fixed charges in thick SiO$_2$, the interface traps between the BP and the SiO$_2$ substrate, and charge transfer from/to neighboring adsorbates on the BP surface. This is commonly observed on 2D devices with an exposed surface in the ambient.[31] The *I-V* characteristics of the Pd contact device, as shown in Figure 2(c) and 2(d), demonstrate inspiring electronic behaviors. Compared to the Ni contact device, the Pd transistor exhibits superior performance in on-state current and mobility. Examining device with the same geometry as above, the Pd contact device has a drain current of 111.8 mA/mm, and extrinsic field-effect mobility of 186.5 $cm^2$/Vs. This larger on-state current of Pd transistor suggests that the Pd/BP contact has a smaller contact resistance compared to the Ni/BP contact. Note that similar threshold voltages ($V_T$) are observed for both devices, where the $V_T$ induced deviation can be fairly ignored in contact resistance comparison. However, if we take a close look at the off-state current at large positive back gate voltage from transfer curves, there is an obvious difference depending on the contact metals. For Ni BP transistors, as the back gate bias sweeps from +26 V to +40 V in Figure 2(b), we clearly observe an increase of drain current, indicating an n-type transistor behavior. As it shows both strong n-type and p-type behaviors in drain current, classify it as an ambipolar transistor, which is rarely seen in previous $MoS_2$ transistors studies. However it is barely seen in high work function Pd contact BP transistors, which only show a monotonic enhancement of p-type drain current with decreasing gate voltage from +40 V to -40 V. Note that the results presented here are also consistent with the previous publication.[25] Due to ultrathin channel thickness and improved electrostatic control, long channel few-layer phosphorene FETs have high current on/off ratio and negligible ambipolar behavior even with Ti or Ni contacts.[25] A detailed examination on Schottky barrier effects on two contacts of BP, which would be discussed in the later parts of this paper, helps us determine the switching mechanism of BP FETs for both p-type and n-type characteristics.

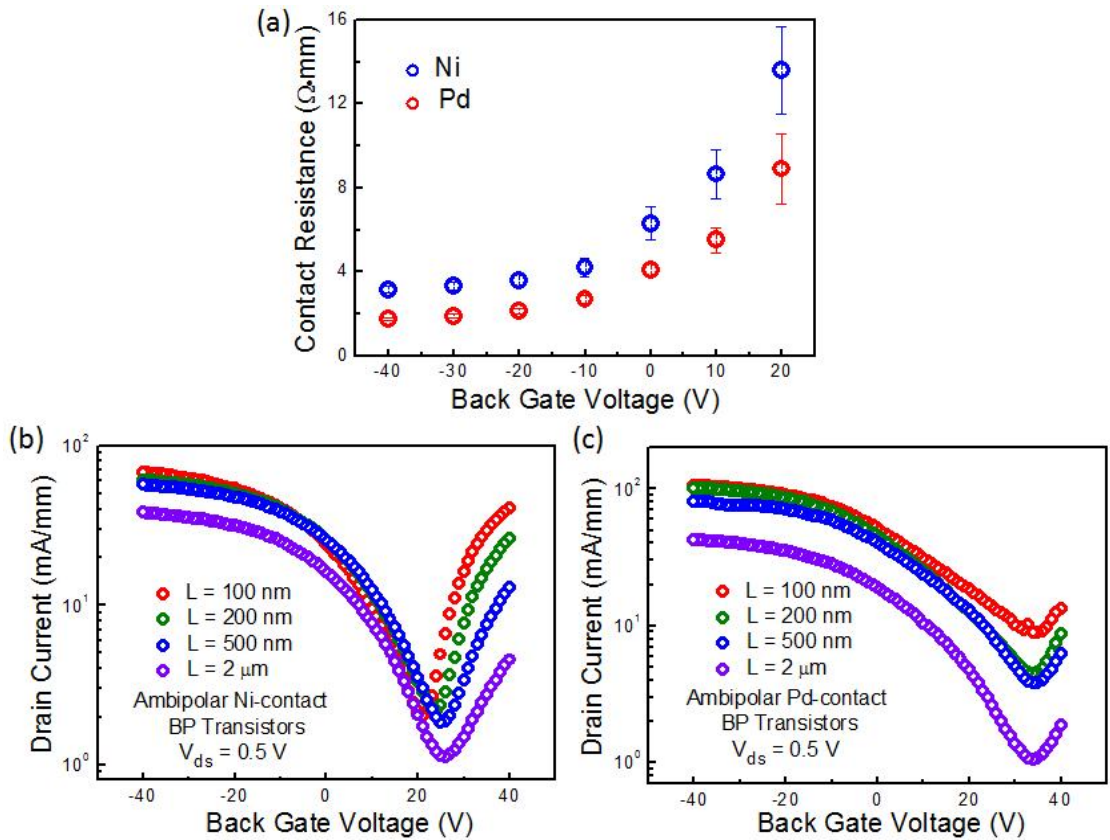

Figure 3: (a) Contact resistance for both Ni and Pd contact metals at various gate biases. Error bars are determined from the standard errors of the linear fitting under different back gate biases. (b) *I-V* transfer characteristic of Ni contact BP FETs with channel length from 2 μm to 100 nm. (c) *I-V* transfer characteristic of Pd contact BP FETs with channel length from 2 μm to 100 nm.

The contact resistance for both Ni and Pd contacts are extracted from TLM structure within long channel regime. Similar to the $MoS_2$ transistors, the contact resistance of BP transistors shows a strong gate dependent behavior as well. The decrease of $R_c$ at lower gate bias is related to the increase of electrostatically doped carrier density in BP under the metal contacts. The higher carrier density induced by the negative back gate voltage enhances the carrier concentration in the BP flake, which leads to a narrower Schottky barrier. Meanwhile, narrowed Schottky barrier would thus facilitate the hole injection from metal into the valence band of BP, leading to a lower contact resistance.[18] To exclude the large absolute errors, high negative bias regions are appropriate to be used for direct comparison of the contact resistance. As measured using TLM structures, the Pd contact resistance is 1.75±0.06 Ω·mm at $V_{bg}$ = -40 V, which is much smaller than the Ni contact resistance of 3.15±0.15 Ω·mm at the same back gate voltage. This nearly 1.8 times reduction in contact resistance is mainly attributed to the different work-function of contact metals. The higher work-function metal Pd with 5.4 eV shows significantly smaller contact resistance than the 5.0 eV metal Ni. Previous $MoS_2$ contact studies had demonstrated that the contact metals were heavily pinned near the conduction band of $MoS_2$ due to its S vacancies at the interface, where both low work function and high work function metals had the similar contact resistances.[18,32] The transfer length ($L_T$) of BP transistors are extracted by the TLM curve[33] and are determined to be 0.72 μm for Pd contact and 1.18 μm for Ni contact at the on-state for 10-20 nm BP channel. We should notice that, the contact width in our experiment is designed as 0.5 μm to maximize the numbers of contact bars on the same BP flake. The designed contact width is a little bit smaller than the calculated transfer length for both metals, indicating a non-full carrier injection from the metals into semiconductor material.[34] This suggests that the extracted contact resistance for both Ni and Pd are overestimated, where the actual contact resistance should be even smaller as contact width could be made larger than the transfer length.

By examining the channel length scaling, we observed clear ambipolar characteristics for both Ni and Pd contacted BP FETs at shorter channel lengths. As depicted in Figure 3(b) and 3(c), enhanced electron current for BP transistors exhibit a strong channel length dependent behavior, where the short channel device with 100 nm channel length demonstrates a better n-type performance compared to the long channel length of 2 μm. A detailed explanation for channel length dependent ambipolar behavior will be provided in the later part of this paper. Notably, contact metals caused n-type performance difference between Ni and Pd is also elucidated in short channel devices. Ambipolar behavior becomes much more pronounced if low work function metals Ni is used for BP FETs. The reason for the more pronounced ambipolar behavior in Ni contact BP FETs is directly attribute to no strong Fermi-level pinning at the metal/BP interface, where the Ni metal Fermi level is closer to the conduction band, resulting in a smaller effective Schottky barrier height compared to high work function metal Pd. In our experiments, total 4 sets of TLM structures, where each set is on the same BF flake, had been carefully fabricated and systematically measured. The results obtained support the conclusion on a conclusion of minimal Fermi-level pinning at metal/BP interface.

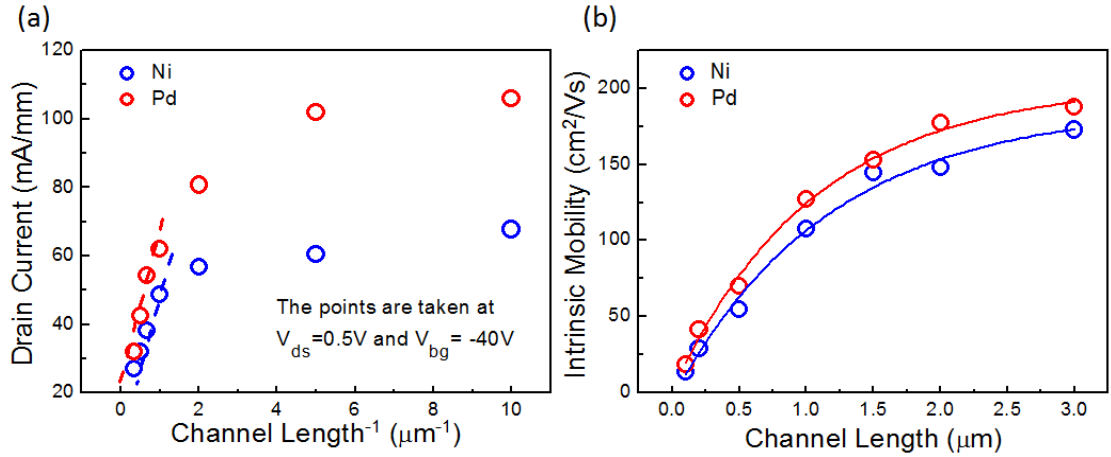

Figure 4: (a) On-state drain current measured at -40 V back gate bias and 0.5 V drain bias on the set of devices fabricated with various channel lengths. The linear dependence property in diffusive region is indicated by the dashed line. (b) Channel length dependent intrinsic field-effect mobility from the set of BP transistors. Ni fitting curve is coincided with Pd fitting curve showing mobility is the fundamental property of BP film itself after extracting contact resistances.

From the classical square-law theory, the drain current is inversely proportional to the channel length $L_{ch}$, where the drain current $I_d$ should be presented as a linear behavior with $1/L_{ch}$. Shown in Figure 4(a), the drain current extracted at $V_{ds}$ = 0.5 V, and $V_{bg}$ = -40 V exhibits a linear relationship at long channel regions. However, starting from 500 nm $L_{ch}$, the drain currents begin to fail the inversely proportional behavior. The Pd contact current becomes to saturate at 105.9 mA/mm at $L_{ch}$ = 100nm, and Ni contact at 67.7 mA/mm as well. Similar to MoS$_2$ transistors, as we scale the BP FET channel length down to 500 nm, the contact resistance starts to be dominated in total resistance. The drain voltage applied on BP FET is mainly dropped at the two contacts, where the contact resistance does not scale with channel length but is present in the device when the contact resistance is comparable to sheet resistance.[18] Next, intrinsic field-effect mobility versus channel length has been studied in BP transistors. The intrinsic carrier mobility, $\mu_{FE}'$ is estimated by $\mu_{FE}' = \mu_{FE}(1-2g_m R_c)^{-1}$. As plotted in Figure 4(b), Ni and Pd are presenting similar intrinsic field-effect mobility, indicating a good uniformity of our ultra-thin BP flake, where both Ni and Pd contacts are fabricated on. Moreover, a decrease of mobility as device channel length shrinks down to 100 nm is realized. This decrement is widely observed at any short channel devices due to the fact that carriers are approaching their saturation velocities at the higher electrical fields.[7,18] The general $v$-E relationship represents $v = \mu E$ and the electrical field in the channel is inversely proportional to channel length. At the low electrical fields, the mobility is constant. However, in the short channel case of very high electrical fields, the velocity starts to approach a value, that is, saturation velocity $\underline{v_s}$ due to the scattering effect and ballistic transport. Therefore, as the velocity reached the saturation value, any further increment of electrical field by scaling the channel length down would result in a decrement of calculated field-effect mobility.[18,34]

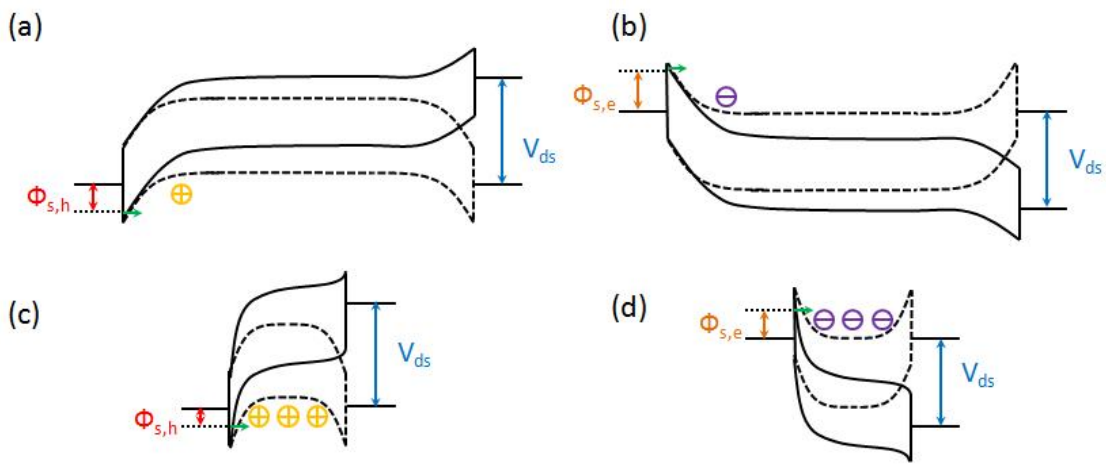

Figure 5: (a) Band diagram for a long channel length BP transistor with p-type behavior. The dashed line stands for the $V_{ds} = 0$ V, and solid line stands for the $V_{ds} \leq 0$ V. (b) Band diagram for a long channel length BP transistors with n-type behavior.

The dashed line stands for the $V_{ds} = 0$ V, and solid line stands for the $V_{ds} \geq 0$ V. (c) Band diagram for a short channel length BP transistor with p-type behavior. The dashed line stands for the $V_{ds} = 0$ V, and solid line stands for the $V_{ds} \leq 0$ V. (d) Band diagram for a short channel length BP transistor with n-type behavior. The dashed line stands for the $V_{ds} = 0$ V, and solid line stands for the $V_{ds} \geq 0$ V.

The nature of Schottky barrier in BP transistors is essential to understand the transistor characteristics. In the previous paragraph, we discussed how the gate bias changes the effective barrier height at the different contacts and hence control the carriers injection at the metal/BP interface. In this part, we further discuss how the drain bias modulates the two Schottky barriers at the contacts, and also the switching mechanism of BP transistors.[34] Similar to $MoS_2$ transistors, there are two asymmetrical Schottky barriers formed at the metal contacts of BP transistors, named as drain barrier and source barrier. However, compared to $MoS_2$ transistors where the drain bias and gate bias can only control the on and off states due to its large bandgap, bias modulations in BP transistors are even more complicated and interesting. Not only the on and off states can be modulated, transistors polarization, n-type and p-type transport behaviors, can also been controlled by the drain bias and gate bias. Thanks to BP's small bandgap, original p-type behavior of BP transistors can be easily turned into n-type transport by controlling channel length, back gate voltages, and drain biases. This kind of transistor characteristics is not easily observed in $MoS_2$ due to its large bandgap and strong contact Fermi-level pinning near conduction band edge.[35]

We firstly define the carriers path for both p-type and n-type BP transistors from the source to drain, that the holes and electrons would encounter the source barrier first, where the carriers would undergo a thermal-assisted tunneling process from the source metal Fermi level to the channel. On the other hand, the carriers in the channel go from valance band or conduction band back to drain metals, for p-type or n-type, respectively. Let us first look at the intrinsic p-type BP transistors with long channel length, as depicted in Figure 5(a). Under large negative gate biases, source and drain contacts start to form the two triangle-shaped Schottky barriers. With further increase of the drain bias, the barrier at the source end remains constant; meanwhile, the drain barrier vanishes, facilitates carriers movement from the source to the drain.[34] To investigate the mechanism of characteristics switch from p-type behavior into n-type, we perform the large positive gate bias for long channel BP transistors, shown in Figure 5(b). Due to the nature of small bandgap structure, carriers can overcome the Schottky barrier at the source and move along in conduction band, showing ambipolar behavior with moderate n-type drain current. With adequate understandings of long channel BP transistors, bias induced switching mechanism on short channel devices remains to be elucidated. When channel length is aggressively scaled down, the drain is close to the source, and the drain bias can start to affect the source barrier, such that the channel carrier concentration is no longer fixed. For long channel devices, a drain bias can change the effective channel length, but the barrier at the source end is independent of the drain bias. However, for a short channel device, the source barrier is no longer fixed. An increase of the drain bias causes the lowering of the source barrier, leading to the injection of extra carriers into the channel,[36] as shown in Figure

5(c). It is similar to drain-induced barrier lowering (DIBL) in short channel Si MOSFETs, where the drain bias decreases the source barrier, thus enhances the off-state current, and makes the device more difficult to turn off.[36] Nevertheless, we need to carefully identify the short channel effects related to BP transistors. In our previous publication, we proposed the drain-induced barrier narrowing (DIBN) model for MoS$_2$ transistors, where the narrowed barrier would reduce Schottky barrier width at source side and enhances the on-state current.[34] Here, we would like to describe the effect as the drain-induced characteristic switch (DICS) in BP field-effect transistors. As shown in Figure 5(d), BP transistors with short channel length under positive back gate bias and drain bias can easily experience its unexpected electrons movement along the conduction band, where the device demonstrates pronounced ambipolar behaviors. Our experimental observations had perfectly matched the proposed DICS effects, which are shown in previous results of Figure 3(b) and 3(c). The scaling of channel length down to 100 nm helps the drain bias start to have impact on the source ends. The lowering of the source barrier would facilitate greater electrons injection into conduction band and the devices would exhibit stronger ambipolar behaviors with enhanced n-type drain currents. The DICS effect on BP transistors will become more obvious as the channel length aggressively scales down to sub-100 nm regime. The geometrical screening length with planar structure is $\lambda = \sqrt{\frac{\varepsilon_s}{\varepsilon_{ox}} t_s t_{ox}}$, where $\lambda$ is the geometrical screening length, $\varepsilon_s$ = 10 is the permittivity of BP,[28,37] $\varepsilon_{ox}$ is the permittivity of gate oxide, $t_s$ and $t_{ox}$ are the thickness of semiconductor channel and gate oxide. The geometrical screening length for this 18.7 nm BP flake is 65.7 nm, which is comparable or only little shorter than our shortest channel length device. In the extreme case when the channel length approaches the screening length, the n-type drain current becomes a strong function of the drain bias, where the n-type behavior may be dominated in BP ambipolar characteristics with a low work function metal, such as Y, La, or Sc. Even with Al metal contacts, our preliminary experimental result shows much pronounced ambipolar effect at larger channel lengths. Our observations in ambipolar BP transistors open a new way to adjust transistors' polarity from p-type to n-type along with the development of doping techniques in 2D materials,[38] and possible realization of future 2D CMOS circuits based on the same channel material BP.

**Conclusion**

In summary, we have studied the BP transistors behaviors with two different high work-function contact metals. 0.6 eV work-function difference between Ni and Pd leads to significantly lower contact resistance on BP using Pd. We explained the switching mechanism of BP transistors, which are controlled by two Schottky barriers at the metal contacts. Moreover, the device characteristics can be alternated from intrinsic p-type into n-type behavior by proper modulation of the channel length, gate bias, and drain bias. With the further development of doping technique and contact engineering, BP transistors show the potential for 2D CMOS logic circuits.

## Methods

Multi-layer BP was exfoliated from the bulk crystal black phosphorus (Smart-elements), and then transferred to a 90 nm $SiO_2$ substrate. All samples were sequentially cleaned by acetone, methanol, and isopropanol to remove the scotch tape residues, and then stored in a nitrogen atmosphere. The thickness of the BP was measured using a Veeco Dimension 3100 atomic force microscope. The Raman optical measurement was conducted in a microscope coupled to a grating spectrometer with a CCD camera. E-beam lithography was used to define the source and drain patterns, using a Vistec VB6. 30/50 nm Ni/Au and 30/70 nm Pd/Au were deposited using e-beam evaporation under $10^{-6}$ Pa pressure, with a deposition rate of 1 Å/s. No annealing was performed after the deposition of the metal contacts. Electrical measurements were carried out with Keithley 4200 semiconductor parameter analyzer and probe station in ambient atmosphere.

**Conflict of Interest**: The authors declare no competing financial interest.

## Acknowledgements

The authors would like to thank Zhe Luo and Xianfan Xu for the optical Raman measurement, and Nathan Conrad for critical reading of the manuscript. This material is based upon work partly supported by NSF under Grant CMMI-1120577 and SRC under Task 2396.

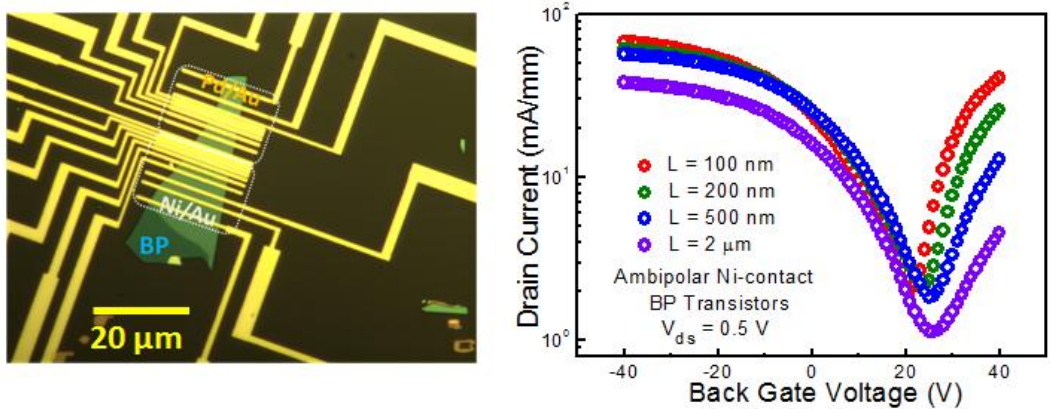

**TOC Graphics**